\documentclass[twocolumn,aps,pra,nofootinbib,groupedaddress,amsmath,amssymb,longbibliography]{revtex4-1}
\usepackage{graphicx}
\usepackage{bm}
\usepackage[colorlinks=true,	linkcolor=blue,urlcolor=blue,anchorcolor=blue,citecolor=blue,bookmarksnumbered]{hyperref}
\usepackage{float}
\usepackage[mathscr]{euscript}
\usepackage[english]{babel}
\usepackage{natbib}

\begin{document}
	\title{Dephasing of ultracold cesium $80D_{5/2}$-Rydberg Electromagnetically Induced Transparency}
	\author{Yuechun Jiao$^{1,3}$}
	\author{Liping Hao$^{1}$}
	\author{Jingxu Bai$^{1}$}
       \author{Jiabei Fan$^{1}$}
	\author{Zhengyang Bai$^{2,3}$}
	\email{zhybai@lps.ecnu.edu.cn}
	\author{Weibin Li$^{4}$}
	\author{Jianming Zhao$^{1,3}$}
	\email{zhaojm@sxu.edu.cn}
	\author{Suotang Jia$^{1,3}$}
	
	\affiliation{$^1$State Key Laboratory of Quantum Optics and Quantum Optics Devices, Institute of Laser Spectroscopy, Shanxi University, Taiyuan 030006, China\\
		$^2$State Key Laboratory of Precision Spectroscopy, East China Normal University, Shanghai 200062, China\\
		$^3$Collaborative Innovation Center of Extreme Optics, Shanxi University, Taiyuan 030006, China\\ 
		$^4$School of Physics and Astronomy, and Centre for the Mathematics and Theoretical Physics of Quantum Non-equilibrium Systems, University of Nottingham, Nottingham, NG7 2RD, UK}

\begin{abstract}
	We study Rydberg electromagnetically induced transparency (EIT) of a cascade three-level atom involving 80$D_{5/2}$ state in a strong interaction regime employing a cesium ultracold cloud. In our experiment, a strong coupling laser couples 6$P_{3/2}$ to 80$D_{5/2}$ transition, while a weak probe, driving 6$S_{1/2}$ to 6$P_{3/2}$ transition, probes the coupling induced EIT signal. At the two-photon resonance, we observe that the EIT transmission decreases slowly with time, which is a signature of interaction induced metastability. The dephasing rate $\gamma_{\rm OD}$ is extracted with optical depth OD = $\gamma_{\rm OD}t$. We find that the optical depth linearly increases with time at onset for a fixed probe incident photon number $R_{\rm in}$ before saturation. The dephasing rate shows a nonlinear dependence on $R_{\rm in}$. The dephasing mechanism is mainly attributed to the strong dipole-dipole interactions, which leads to state transfer from $nD_{5/2}$ to other Rydberg states. We demonstrate that the typical transfer time $\tau_{0(80D)}$ obtained by the state selective field ionization technique is comparable with the decay time of EIT transmission $\tau_{0({\rm EIT})}$. The presented experiment provides a useful tool for investigating the strong nonlinear optical effects and metastable state in Rydberg many-body systems.
\end{abstract}
	
	\maketitle

%The optical depth (OD) of the medium on EIT resonance is defined to characterize the EIT decay.
%\keywords{cesium nD Rydberg state, electromagnetically induced transparency, nonlinear}
%\pacs{32.80.Rm, 42.50.Gy, 42.65.Sf}
%\maketitle it attracts great attentions in the communities of nonlinear optics~\cite{Pritchard2010,Sevincli2011},
%quantum information~\cite{Quantum_Saffman_2010}, and many-body physics~\cite{browaeys_many-body_2020}.The interactions
%are flexibility tailored by altering appropriate Rydberg states and by applying external fields.

\section{Introduction}

Due to the strong interaction ($\propto n^{11}$ with $n$ principal quantum number)~\cite{Gallagher1994}, Rydberg atoms provides an ideal platform to implement quantum information and quantum simulation~\cite{Jaksch2000,Viscor2015,Isenhower2010a,Georgescu2014} and investigate interaction induced cooperative optical nonlinearities~\cite{Pritchard2010}.  The optical nonlinear effects are induced by Rydberg atom interactions~\cite{Parigi2012,Peyronel2012a}, i.e., van der Waals (vdW) interactions~\cite{Singer2005,Vermersch2015,Jiao2016e} and dipole-dipole interactions~\cite{Vogt2006, Comparat2010}. Strong atomic interactions can be effectively mapped onto photon-photon interactions via electromagnetically induced transparency (EIT) ~\cite{Pritchard2010}.  Due to the cooperative effects, the optical nonlinearity can be greatly enhanced~\cite{Pritchard2010, Sevincli2011, bai_enhanced_2016, Two_Chen_2021}. Based on this, Rydberg EIT experiments are employed to measure a radio-frequency electric field~\cite{Sedlacek2012} with a room-temperature cell, and to realize few-photon optical nonlinearities~\cite{Dudin2012,Baur2014,Li2013,Tiarks2014} with an ultracold sample, such as the efficient single photon generation~\cite{Dudin2012}, entanglement generation between light and atomic excitations~\cite{Li2013}, single-photon switches~\cite{Baur2014,Two_Chen_2021,Ding_2022_Facilitation} and transistors~\cite{Gorniaczyk2014,Tiarks2014,gorniaczyk_enhancement_2016}. The resonant dipole-dipole interaction between two individual Rydberg atoms~\cite{Ravets2015} is angular dependent. Recently, the anisotropic Rydberg interaction have been adopted to investigate Rydberg polaritons by using  Rydberg $nD$-state~\cite{Tresp2015}.

%Rydberg electromagnetically induced transparency (EIT)~\cite{Mohapatra2007a}, applying for a non-destructive detection of Rydberg atom, provides a coherent mechanism to enhance optical nonlinear effects~\cite{Pritchard2010, Sevincli2011, bai_enhanced_2016, Two_Chen_2021}.

%The EIT also can be used to map the interaction between Rydberg atoms onto a strong optical transition~\cite{Pritchard2010}.

In this work, we present a Rydberg EIT spectrum of a cascade three-level cesium atom involving  $80D_{5/2}$ state in a dipole trap. Under the two-photon resonance condition, we observe that EIT transmission displays a slow decrease with time. This could be a signature of interaction induced metastability~\cite{Towards_Macieszczak_2016,Metastabl_Macieszczak_2017}. Due to the large dipole matrix elements, the cesium $nD$-state atom has strong dipole interactions with energetically close $(n+1)P$ state.
We find that, strong dipole interactions  between the $nD$ and $(n+1)P$ state leads to a fast decay of $nD$ to $(n+1)P$ state, and the dephasing of the transmission spectrum. A theoretical model is built to understand EIT dephasing mechanism. The nonlinear dependence on dephasing rate has also been investigated.

The remainder of the article is arranged as follows. In Sec.~\ref{sec_exp}, we introduce our experimental setup. In Sec.~\ref{sec_dephasing}, we reveal Rydberg EIT spectrum and its dephasing experimentally. In Sec.~\ref{sec_model}, we present a simple theoretical model to reveal EIT dephasing mechanism. In Sec.~\ref{state_transfer}, we measure the fast decay process on $80D_{5/2}$ state. Finally,  we summarize the main results obtained in this work.

\section{Experimental Setup}\label{sec_exp}

\begin{figure}[ht]
\centering
\includegraphics[width=0.5\textwidth]{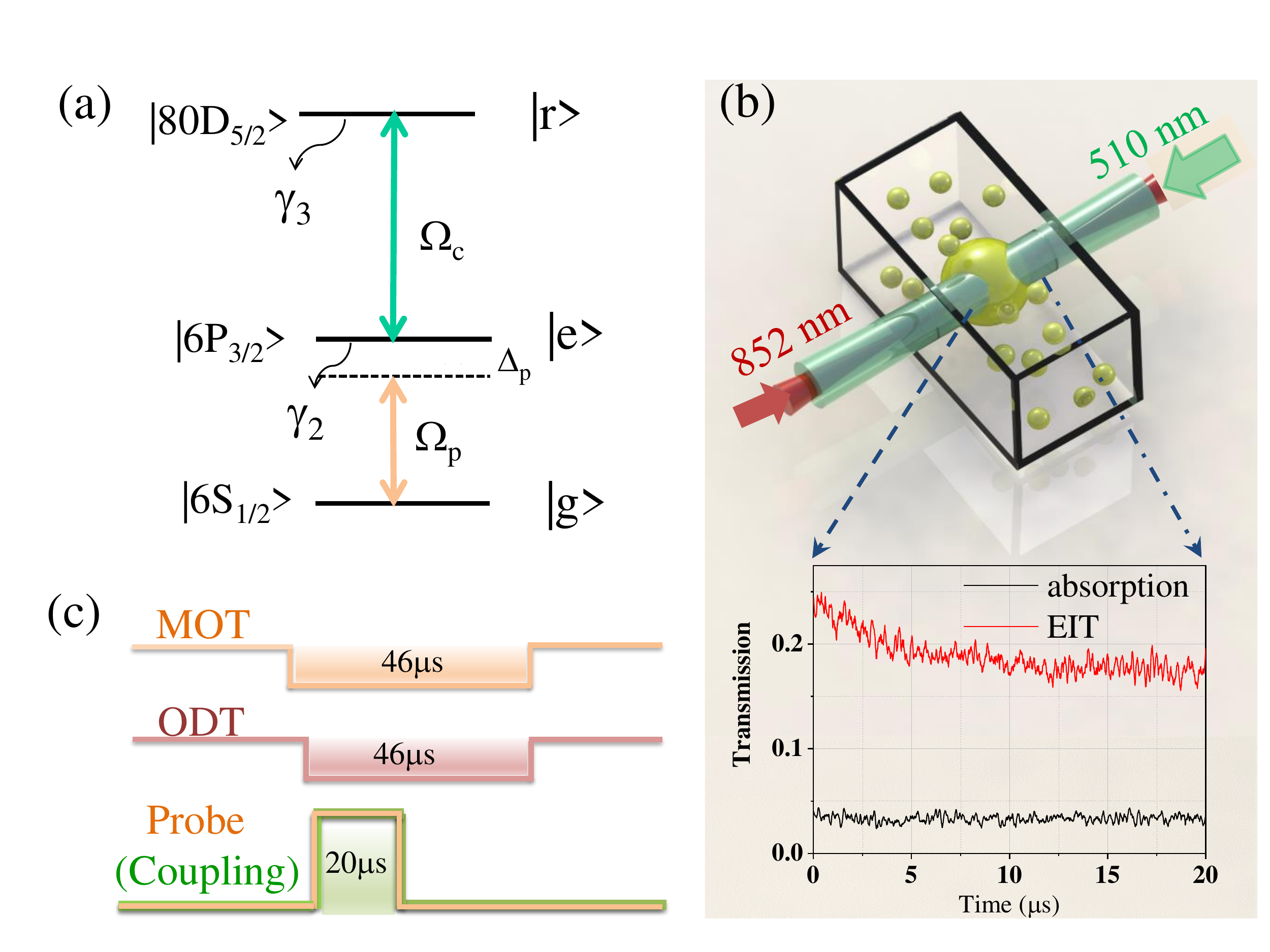}
\caption{(color online) (a) Atomic level scheme. A weak probe laser field with Rabi frequency $\Omega_{p}$ drives the lower transition, $|g\rangle = |6 S_{1/2}, F=4\rangle $  $\to$
$|e\rangle=|6P_{3/2}, F'=5\rangle $. The strong coupling laser  (Rabi frequency $\Omega_c$) couples the transition $|e\rangle$ $\to$ $|r\rangle = |80D_{5/2}\rangle$. 
%Due to dipolar interaction, atoms in state $|80D_{5/2} \rangle $ can decay to auxiliary Rydberg state $|r^\prime\rangle$. 
(b) Experiment setup. The
coupling and probe beams are counter-propagated through the MOT center and overlap with the dipole trap
beam. The transmission of probe beam is detected with a single photon counting module (SPCM). The inset shows EIT dephasing behaviors (red) and probe absorption (black; without coupling beam). The data are taken by the sum of 3000 experimental cycles. 
%The inset shows Rydberg polariton induced by the blockade effect. In blockade region
%with the blockade radii $r_B$, the excitation of one Rydberg atom shifts the level of neighbor atoms that were considered as two-level atoms and absorb/scattering probe photons.
(c) Experimental timing. After switching off the MOT and dipole trap beams, Rydberg-EIT coupling and probe lasers are turned on for 20$~\mu$s, during which the probe laser frequency is ramped through the $|g\rangle$ $\to$ $|e\rangle$ transition over $\pm$15~MHz by a double-passed AOM.}
\end{figure}

Our experiment is performed in a cesium magneto-optical trap (MOT) with optical dipole trap (ODT).The beam waist of dipole trap is 45 $\mu$m. A schematic of relevant levels and the experimental setting are shown in Fig.~1 (a) and (b).
A three-level system, shown in Fig.~1(a), consists of a ground state $|6S_{1/2}, F = 4\rangle$ ($|g\rangle$), intermediate state $|6P_{3/2},  F' = 5 \rangle$ ($|e\rangle$) and  Rydberg state $|80D_{5/2}\rangle$ ($|r\rangle$). A weak probe beam (Rabi frequency $\Omega_p$, 852-nm laser with a 100-kHz linewidth), provided by an external cavity diode laser (Toptica, DLpro), drives a lower transition and the frequency is stabilized to the $|6S_{1/2}, F = 4\rangle$ $\to$ $|6P_{3/2},  F' = 5 \rangle$ transition using the polarization spectroscopy method. The coupling beam (Rabi frequency $\Omega_c$) provided by a commercial laser (Toptica, TA-SHG110) with linewidth 1~MHz drives Rydberg transition, $|6P_{3/2}, F'=5 \rangle$ $\to$ $|80D_{5/2}\rangle$.
%The coupling (probe) beam has a Gaussian radius in the MOT center 30~$\mu$m (9~$\mu$m) and overlaps with the dipole trap beam. 
The coupling laser frequency is stabilized to the Rydberg transition using a Rydberg EIT reference signal obtained from a cesium room-temperature vapor cell~\cite{Jiao2016f}. The weak probe laser and strong coupling laser, with respective Gaussian radius of 9~$\mu$m and 30~$\mu$m, are overlapped and counter-propagated through the MOT center, see Fig.~1(b). The probe laser is scanned using a double-passed acousto-optic modulator (AOM) that covers the lower transition.
The transmission of the probe laser, Rydberg EIT spectrum, is detected with a single photon counting module (SPCM) and processed with \textit{Labview} program. The glass MOT is surrounded by three pairs of field-compensation Helmholtz coils, which allow us to reduce stray magnetic fields via EIT Zeeman splitting, corresponding stray field less than
5~mG. In our experiment, the peak density of atomic cloud about 10$^{11}$~cm$^{-3}$ is measured by shadow imaging and the temperature of atomic cloud is about 100 $\mu$K. The estimated Rydberg density is 2.4 $\times$ 10$^{8}$~cm$^{-3}$.

The experimental timing is shown in Fig.~1(c) with the whole time 200 ms, corresponding to a repetition rate of 5 Hz. In each cycle, after turning off the trap beams, we switch on the coupling and probe lasers for 20 $\mu$s, during which the probe-laser
frequency is swept across the $|6S_{1/2}, F = 4\rangle$ $\to$ $|6P_{3/2},  F' = 5 \rangle$ transition, meanwhile the  power is fixed using a proportional-integral-derivative controller (PID) feedback loop that controls the radio-frequency power supplied to the 852-nm AOM. The data shown in Fig.~1(b) is taken by the sum of 3000 cycles.

\section{Experimental observation of Rydberg EIT and transmission dephasing}\label{sec_dephasing}

\begin{figure}[ht]
\centering
\includegraphics[width=0.5\textwidth]{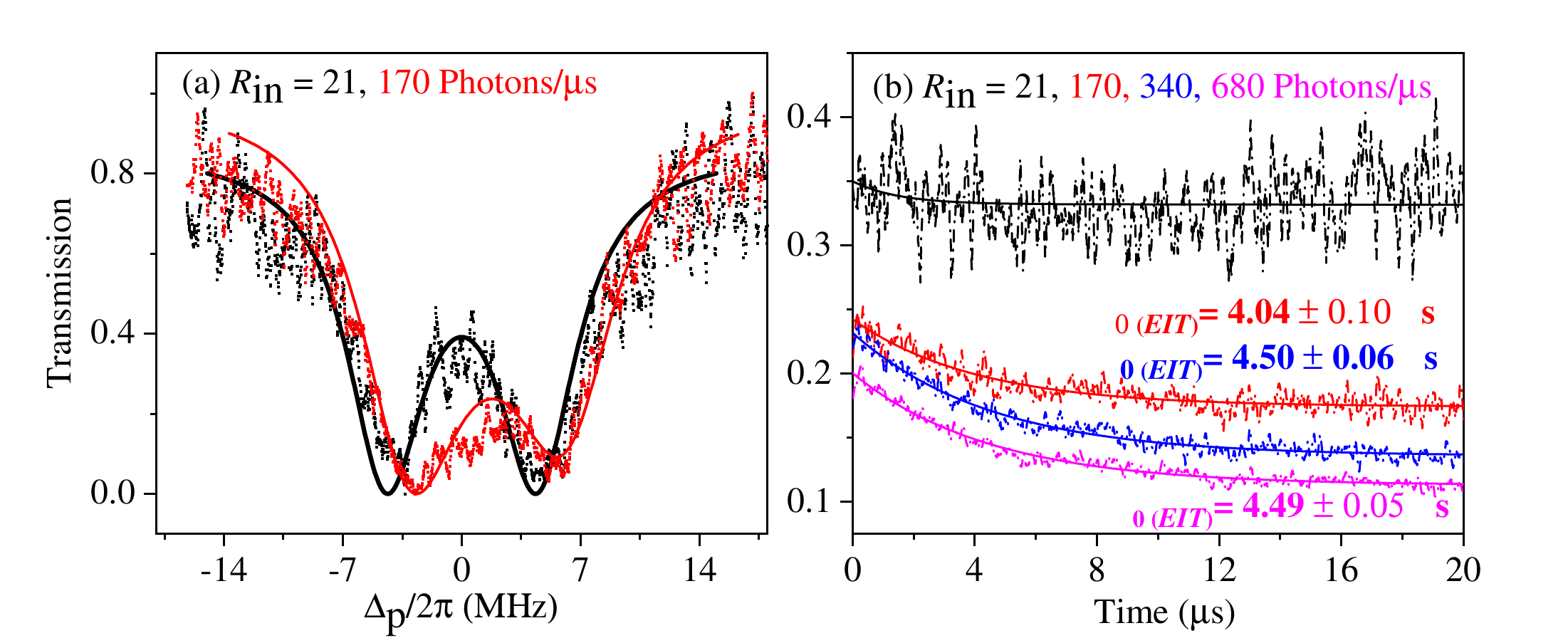}
\caption{(color online) (a) Rydberg-EIT spectra with a coupling laser, $\Omega_c /2\pi$ =10.6~MHz, resonant with the $|6P_{3/2}, F'= 5\rangle$ $\to$ $|80D_{5/2}\rangle$ transition, and a probe frequency scanning across the lower transition, $|6S_{1/2}, F = 4\rangle$ $\to$ $|6P_{3/2},  F' = 5 \rangle$, at a probe-photon rate $R_{\rm in}$ = 21 photons/$\mu$s and 170 photons/$\mu$s, respectively. The solid lines are the fittings using the density matrix equation of a three-level atom. The EIT transmission displays a strong suppression and blue shift with increasing probe $R_{\rm in}$. (b) EIT transmissions for the indicated probe photon input rates at two-photon resonance condition. The transmission remains same for $R_{\rm in}$ = 21 photons/$\mu$s, whereas decrease with $R_{\rm in}$ at the beginning and then decay with the EIT maintain time for larger $R_{\rm in}$ cases. The solid lines denote the exponential fittings. For higher $R_{\rm in}$ the characteristic time are $\tau_{0({\rm EIT})}$ = 4.04 $\pm$ 0.10~$\mu$s, 4.50 $\pm$ 0.06~$\mu$s and 4.49 $\pm$ 0.05 $\mu$s, respectively. }
\end{figure}

Due to the quantum interference effect~\cite{fleischhauer_electromagnetically_2005}, the probe transmission $T$ increases when the coupling and probe laser frequencies satisfy the two-photon resonance [see Fig.~1(a)].
Using the rotating wave approximation and in the interaction picture~\cite{Hao2018}, the eigenstate of three-level Hamiltonian can be obtained with  $|D\rangle \propto \Omega_c^{*} |g\rangle - \Omega_p |r\rangle$ at two-photon resonance.  For conventional EIT, the system works in dark state and
the probe pulse suffers little optical absorption. For the three-level scheme, all atoms are  initially prepared in state $|g\rangle$.  The EIT system can evolve to the dark state  $|D\rangle$ with time $1/\gamma_2$. However, as shown in Fig.~1(b), in $nD_{5/2}$-Rydberg EIT system, the EIT transmission slowly decreases with time. Interesting, the time scale of the dephasing is much longer than the lifetime of intermediate state $|e\rangle$. The dipolar interaction can lead to many-body dephasing~\cite{Tresp2015, Dong_Cold_2020} in case of Rydberg $nD$ state.
%State $|D\rangle$ does not overlap with the fast decaying intermediate state $|e\rangle$ and is named as the dark state.

%due to Rydberg atom state transfer of  $|80D_{5/2} \rangle $  $\to$ $|81P_{3/2}\rangle$,
%see .
%However, in the strong-interaction regime,when a probe photon is transmitted through the sample without absorption, all other atoms act as two-level atoms due to the
%coupling frequency detuned by the Rydberg-atom pair interaction, and absorb/scatter the excess photons, see Fig.~1(b), which yields the decay and dephasing of the EIT transmission.

We present EIT spectra with 80$D_{5/2}$ Rydberg state in Fig.~2(a). The probe field is adopted as incident photon rates $R_{\rm in}$ = 21 photons/$\mu$s (black dashed line) and 170 photons/$\mu$s (red dashed line) with $\Omega_c$ = $2\pi\times10.6$~MHz.
It is found that the transmission decreases with increasing $R_{\rm in}$ and accompanies with the EIT-peak shift. The EIT transmission rate is around 40\% for $R_{\rm in}$ = 21 photons/$\mu$s, and decreases to 20\% when $R_{\rm in}$ increases to 170 photons/$\mu$s. We attribute the reduction of optical transmission to atomic interactions between Rydberg states.
Besides, when $R_{\rm in}$ = 170 photons/$\mu$s, the EIT peak has a blue shift about $2.5$~MHz compared with the case for $R_{\rm in}$ = 21 photons/$\mu$s.
For Rydberg EIT, dark-state polariton is very sensitive to other Rydberg excitation in the strong interaction regime. This is because when two Rydberg polaritons propagate inside the medium with a distance $r$, they experience an interaction induced energy shift $U(r)$.
It gives rise to a non-vanishing Im$(\chi)$ of probe beam where $\chi$ optical susceptibility of system and therefore leads to strong absorption and a shift on EIT spectra~\cite{Pritchard2010, Sevincli2011, bai_enhanced_2016}.

%\section{Dephasing of Rydberg EIT}

To further investigate the dephasing feature for $nD$ state, we vary the probe-photon rate $R_{\rm in}$ with fixed coupling field $\Omega_c /2\pi$ =10.6~MHz. Both the coupling and probe lasers frequencies are on resonance. As shown in Fig.~2(b), the time dependence of the transmission is plotted with different $R_{\rm in}$.
For low probe photon rates (i.e., $R_{\rm in}$ = 21 photons/$\mu$s), the transmission is almost a constant, but when increasing $R_{\rm in}$, transmission exhibits a slow decrease with time $t$. By fitting the data in panel (b) with the exponential function $T = A \exp(-t / \tau_{0({\rm EIT})}) + T_0$, the decay time $\tau_{0({\rm EIT})}$ = 4.04 $\pm$ 0.10~${\rm \mu s}$, 4.50 $\pm$ 0.06~${\rm \mu s}$ and 4.49 $\pm$ 0.05~${\rm \mu s}$ can be extracted for different $R_{\rm in}$. The decay time is much longer than the lifetime in state $|e\rangle$ (i.e., $1/\gamma_2\sim0.03{\rm \mu s}$).

%We attribute this decay effect to the dephasing of Rydberg EIT system that is induced by the dipole-dipole interaction between Rydberg atoms.

%$R_{\rm in}$ and observe a decay of EIT transmission over probe duration time, while keeping the coupling and probe frequencies are fixed and two photon detuning $\delta$ = 0.

\begin{figure}[ht]
\centering
\includegraphics[width=0.5\textwidth]{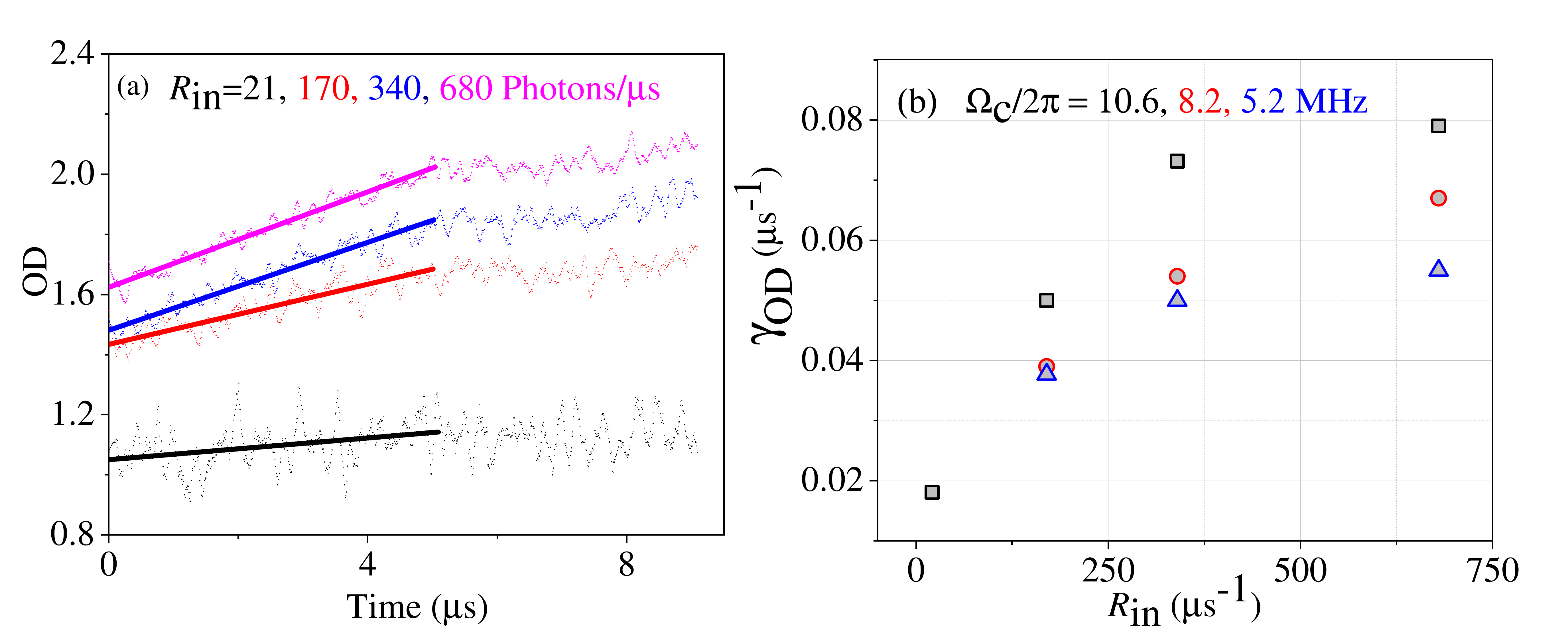}
\caption{(color online) (a) The optical depth of the 80$D_{5/2}$ EIT transmission taken the logarithm of spectra in Fig.~2(b) for indicated photon incidence rates $R_{\rm in}$ at fixed $\Omega_c$ = 2$\pi \times $10.6~MHz. The solid lines are linear fittings before t = 5 $\mu$s to the data to extract the dephasing rates $\gamma_{\rm OD}$. (b) The dephasing rates $\gamma_{\rm OD}$ as a function of $R_{\rm in}$ for coupling Rabi frequency $\Omega_c/2\pi$ = 10.6, 8.2 and 5.2~MHz, respectively.}
\end{figure}

In order to reveal the dephasing mechanism, we define the effective optical depth (OD) of the medium as the logarithm of transmission [i.e., OD = -ln($T$)]~\cite{Tresp2015}.
%EIT$_{trans}$, that is, OD = log(EIT$_{trans}$).
The time evolution of OD is shown in Fig.~3(a) [corresponding to the results in Fig.~2(b)].
One sees that OD approximately linearly increases with time before t = 5 $\mu$s. By neglecting saturation effects, we can
redefine $OD = \gamma_{\rm OD}t$ where $\gamma_{\rm OD}$ reflects creation rate of optical
density by decoupled impurities~\cite{Tresp2015}.
At small probe photons (i.e., $R_{\rm in} \lesssim $ 340~$\mu$s$^{-1}$), the extracted rate $\gamma_{\rm OD}$ displays a linear increase with $R_{\rm in}$ and then saturates for large $R_{\rm in}$ [see Fig.~3(b)]. This is because the system is not fully in a blockade regime at small probe photons. Therefore, with increasing photon number, more Rydberg atoms are excited, which increase the dipole-dipole interaction, leading to increased $\gamma_{\rm OD}$. However, after the system enter fully blockade regime, we can't excite more Rydberg atoms so that the dephasing rate $\gamma_{\rm OD}$ shows saturation. This is a signature of many-body dephasing. 
We calculate the group velocity of probe photon is around $3920$~m/s under our experiment condition with $\Omega_c=10.6$ MHz. By considering the length of our atomic cloud $1$ mm, the propagation time of photon through the cloud is around $0.26$~${\rm \mu s}$. The Rydberg blockade radius is about 10~${\rm \mu m}$ for 80D$_{5/2}$. Under this condition, almost $100$ atoms can be excited to Rydberg state. Thus we can calculate that in one microsecond, the maximum number of Rydberg excitation is around 385, where $100\,{\rm atom}/0.26$\,${\rm \mu s}$\,$\simeq385\,/{\rm \mu s}$. Thus, the critical value for $R_{\rm in}$ is around 385~/${\rm \mu s}$. When the photon incidence rates $R_{\rm in}$ is smaller than 385~/${\rm \mu s}$, the system is not in the blockade regime. When increasing the intensity of the  probe laser, the system can works in the full blockade regime. This estimation is agreed with the dephasing rates $\gamma_{\rm OD}$ in Fig.~3(b). We also change $\Omega_c$ and measure  EIT dephasing rates $\gamma_{\rm OD}$ versus $R_{\rm in}$. The results show a similar nonlinear dependence on $R_{\rm in}$ [see Fig.~3(b)].

%of the EIT transmission that takes the logarithm of spectra in Fig.~2(b).

%Below we discuss the dephasing mechanism causing the decay of EIT transmission rate. For the cascade three-level atom discussed here, a few dephasing mechanisms may result in a
%decay of EIT transmission.

%\subsection{Stray field induced dephasing}
%We consider firstly the decay due to the level shifts induced by the spatially inhomogeneous electric or magnetic field. The stray magnetic field in the MOT center is compensated
%with three Helmholtz coils, a $\lesssim$ 5~mG magnetic field is estimated by the EIT Zeeman splitting. A stray electric field in the MOT center is measured to less than 20~mV/cm
%with Stark spectra of 80$D_{5/2}$. These conditions yield the level shift less than 1~MHz and a negligible EIT dephase rate.

\section{Analysis of EIT dephasing mechanism}\label{sec_model}
The vdW interaction between $nD_{5/2}$ pair leads to energy level shifts and dipole interaction of $nD_{5/2}$ and nearest Rydberg states yield state transfer of $nD_{5/2}$ to other Rydberg states. As no microwave field is present in the experiment,   the other states are excited due to the spontaneous decay from $nD_{5/2}$ state. We have found~\cite{Hao_Observation_2021} that $nD$-$(n+1)P$ transition is the strongest in our experiment conditions (see Sec. 5 for details). Hence this leads to a two stage processes. Rydberg $nD_{5/2}$ state will decay to $(n+1)P_{3/2}$ through spontaneous decay. The dephasing can then be induced in that regime where dipole-dipoles interactions couple nearly degenerate Rydberg pair states~\cite{Tresp2015, Dong_Cold_2020}. A full model to describe the dephasing is rather complicated. In this section, we will focus on the dephasing effects with a simplified model, which nonetheless captures the main effects.

% Here we refer all auxiliary  states as the removed state  $|r^\prime\rangle$.

Considering the three-level scheme in Fig.~1(a), the dephasing rate of Rydberg state $\gamma_{3}$ is around $\gamma_{r}$ + $\Gamma_{re}$ $\thickapprox \gamma_{r}$. 
Rydberg atom has long lifetimes ($1/\Gamma_{re} \sim n^3$) on the order of 100~$\mu$s. $\gamma_{r}$ represents the dephasing of the atomic coherence (originated from atomic collisions, residue Doppler effect, dipole-dipole interaction between the Rydberg atoms, finite laser linewidth). The dephasing rate of the intermediate state $|e\rangle$ denotes $\gamma_{2}$ = $\gamma_{e}$ + $\Gamma_{eg}$ with spontaneous decay rate $\Gamma_{eg}$ and interaction induced decay $\gamma_{e}$. In our physical system, $\gamma_{e}$ is much smaller than $\Gamma_{eg}\simeq 2\pi \times 5.2$~MHz, thus $\gamma_{2}$ $\thickapprox \Gamma_{eg}$ .
The collective dissipation can emerge in dense atomic gases, typically through two-body dipolar couplings~\cite{Dong_Cold_2020}.
Here we adopt the effective dephasing $\gamma_3^{\rm eff}$  (i.e., $\gamma_{r}=\gamma_3^{\rm eff}$), and seek the relation between the transmission and dipolar interaction induced dephasing.%Here, we adopt the effective dephasing $\gamma_3^{\rm eff}$ to replace the depopulation process (i.e., $\gamma_{r}=\gamma_3^{\rm eff}$), and seek the relation between the transmission and dipolar interaction induced dephasing. By eliminating the auxiliary Rydberg level, 

The dynamics of the effective three-level system can be modeled by the quantum master equation for the many-atom density operator $\rho$:
\begin{eqnarray} \label{master}
\dot{\rho}=-i[\hat{H}_{\rm eff},\rho]+D_1(\rho)+D_{\rm eff}(\rho),
\end{eqnarray}
The effective Hamiltonian in the equation is given by
%
%\begin{subequations}\label{Hamiltonian1}
\begin{eqnarray}\label{Hamiltonian1}
\hat{H}_{\rm eff} =&&
\sum_{j=1}^{N}\left[-\Delta_p\hat{\sigma}_{ee}^j(\mathbf{r},t)
-(\Delta_p+\Delta_c)\hat{\sigma}_{rr}^j(\mathbf{r},t)+\frac{\Omega_p}{2}\hat{\sigma}_{eg}^j(\mathbf{r},t)\right.\nonumber\\
&&\left.+\frac{\Omega_c}{2}\hat{\sigma}_{re}^j(\mathbf{r},t)+\sum_{k\neq j}^N\frac{V_{jk}}{2}\hat{\sigma}_{rr}^j\hat{\sigma}_{rr}^k + {\rm H.c.}\right],
\end{eqnarray}
%\end{subequations}
%
with $\hat{\sigma}_{ab}(z_j)\equiv|a_j\rangle\langle b_{j}|$\,  ($z_j$ is the position of $j$th atom in the respective ensemble) and H.c. representing Hermitian conjugate of the preceding terms. $V_{jk}=C_6/|{\bf r}_j-{\bf r}_k|^6$ is the vdW potential with the dispersive coefficient $C_6\propto n^{11}$.
The dissipative effects are described by the Lindblad form $D_1(\rho)$,
%
%\begin{subequations}\label{Lindblad1}
\begin{eqnarray}\label{Lindblad1}
%\begin{aligned}
D_1(\rho)=&&\sum_{j=1}^{N}\Gamma_{eg}\left(\hat{\sigma}_{ge}^j\rho\hat{\sigma}_{eg}^j-\frac{1}{2}\{\hat{\sigma}_{ee}^j, \rho\}\right),
\end{eqnarray}
%\end{subequations}
where $D_1(\rho)$ denotes the decay from state $|e\rangle$ to $|g\rangle$. The effective dephasing term $D_{\rm eff}(\rho)$ is introduced,
%\begin{subequations}\label{dephasing}
\begin{eqnarray}\label{dephasing}
%\begin{aligned}
D_{\rm eff}(\rho)=&&\sum_{j=1}^{N}\gamma_3^{\rm eff}\left(\hat{\sigma}_{33}^j\rho\hat{\sigma}_{33}^j-\frac{1}{2}\{\hat{\sigma}_{33}^j, \rho\}\right),
\end{eqnarray}
%\end{subequations}
%
Due to the dephasing and spectral shift of the transparency resonance [see Fig.2(a)], we employ the theoretical description of
individual atoms coupled to a mean field (MF) to analyze the EIT spectrum with strong Rydberg interactions~\cite{Schempp_Coherent_2010, Pritchard_Cooperative_2010, Spectral_Han_2016}. 
In the MF approximation, the many-body density matrix $\rho$ is decoupled into individual ones through $\hat{\rho} \approx \Pi_i$ $\hat{\rho}_i$. In the thermodynamic limit,  the optical Bloch equations for the three-level system can be obtained, where elements of the density matrix are represented with $\rho_{ab}=N^{-1}\sum_j\langle\hat{\sigma}_{ab}^j\rangle$.
As indicated in our experiment, over a long time evolution, the interactions between Rydberg state leads to a MF shift where $\Delta_c\rightarrow\Delta_c+V\rho_{rr}$ with MF interaction energy $V=N^{-1}\sum_{k\neq j}V_{jk}$.

\begin{figure}[ht]
\centering
\includegraphics[width=0.5\textwidth]{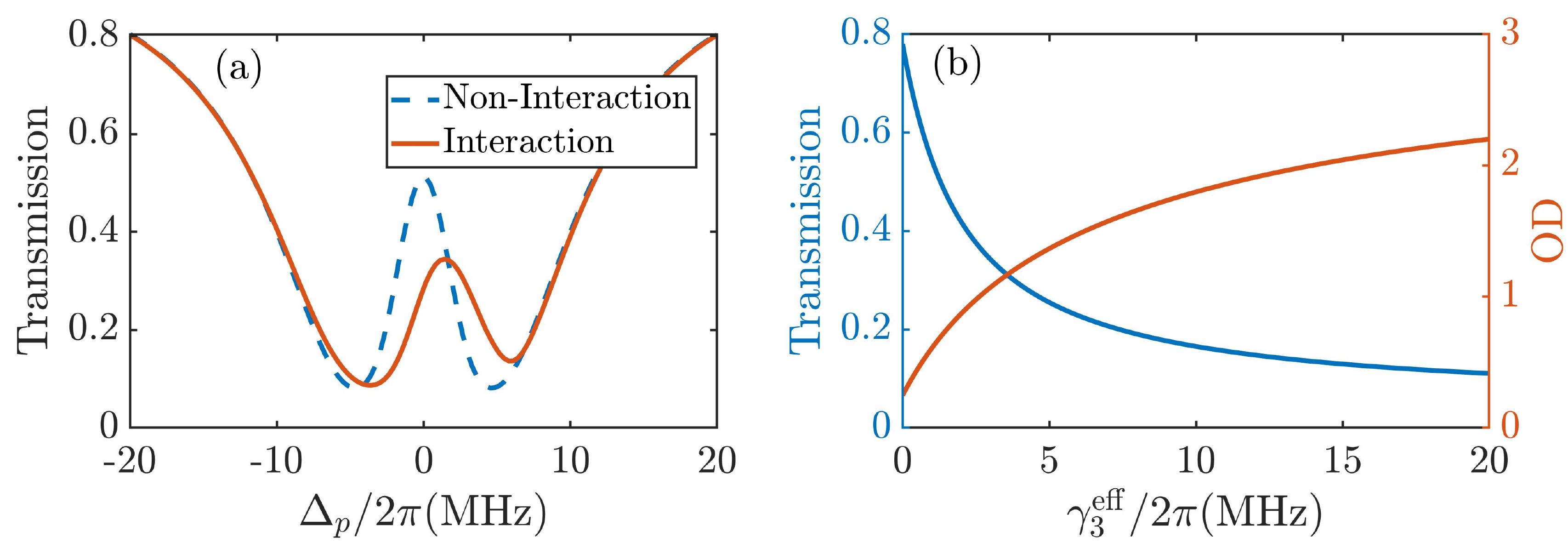}
\caption{(color online) (a) Theoretical results of EIT transmission varies with $\Delta_p$ for $V=0$ and  $V\neq0$. (b)  EIT transmission and corresponding OD versus effective dephasing rate $\gamma_3^{\rm eff}$ of Rydberg state that accounts for the interaction between Rydberg atoms with $\Omega_p /2\pi$ =1.04~MHz, $\Omega_c /2\pi$ =10.6~MHz and atomic density 1$\times$10$^{11}$cm$^{-3}$.}
\end{figure}

\begin{figure}[ht]
\centering
\includegraphics[width=0.5\textwidth]{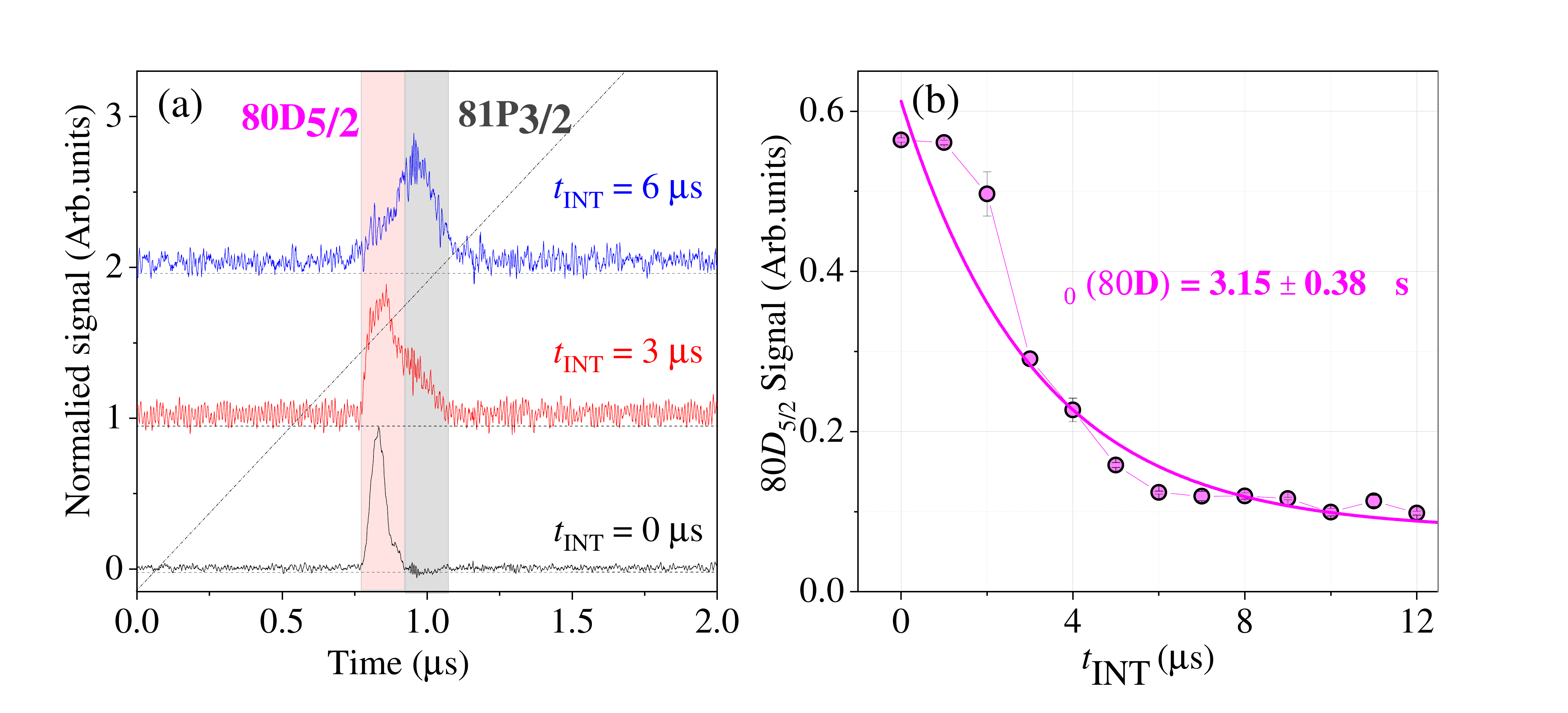}
\caption{(color online) (a) Normalized time of flight (TOF) signals for laser excitation to 80$D_{5/2}$ state with indicated interaction times, $t_{\rm INT}$. The three traces are set vertically offset for clarity, with the respective zero levels shown as horizontal dashed lines. The gates for the 80$D_{5/2}$ and 81$P_{3/2}$ states signals are shown as a light pink and gray shaded regions, respectively. (b) Measurements of the 80$D_{5/2}$ state as a function of the interaction time $t_{\rm INT}$. Initial prepared 80$D_{5/2}$ Rydberg atoms transfer to nearby Rydberg states $|r^\prime\rangle$ during $t_{\rm INT}$ due to the strong resonant dipole interaction. The solid line shows the exponential fitting with the characteristic time $\tau_{0(80D)}$ = 3.15 $\pm$ 0.38 $\mu$s.}
\end{figure}

By numerically solving the MF Bloch equation, one can obtain EIT transmission varies with $\Delta_p$. As shown in Fig. 4(a), there is an increasing blue shift of transparency away from the non-interaction EIT resonance. This is because that the shifted Rydberg state detunes the EIT windows. The similar EIT signal is also observed experimentally [see Fig.2(a)].
To obtain the dependence of the EIT transmission on the effective dephasing induced by Rydberg interactions, we calculate EIT spectra for a series of effective $\gamma_3^{\rm eff}$, accounting for the many-body dephasing by Rydberg atoms. In Fig.~4(b), the EIT transmission as a function of $\gamma_3^{\rm eff}$ is plotted. The system works at two-photon detuning $\delta$ = 0 with the probe
$\Omega_p /2\pi$ =1.04~MHz, $\Omega_c /2\pi$ =10.6~MHz and atomic density $N_a=1\times10^{11}$ cm$^{-3}$. It shows that the EIT transmission decrease with $\gamma_3^{\rm eff}$. The EIT transmission decreases to 50\% when $\gamma_3^{\rm eff}=2\pi\times1.5$~MHz. It is consistent with the trend of our experimental data [see Fig.~2(a)]. For comparing, we also plot the corresponding OD in Fig.~4(b), as expected, calculated OD of the probe beam increase as $\gamma_3^{\rm eff}$. We should note that the presented theoretical model is based on MF equation by simply varying the effective decay rate $\gamma_3^{\rm eff}$. Beyond the present model, many-body quantum model need to be developed to gain better understanding of the metastable dynamics~\cite{Towards_Macieszczak_2016, Metastabl_Macieszczak_2017}. We will discuss it somewhere else.

\section{Test of the decay in $80D_{5/2}$ state}\label{state_transfer}
The dephasing effects arise from many physical reasons, e.g., atomic collisions, residue Doppler effect, depopulation
between the Rydberg atoms, or finite laser linewidth. To test the collective dephasing process, we also conduct the experiment to measure the fast population transfer from $80D_{5/2}$ to $81P_{3/2}$. For 80$D_{5/2}$ state used in this work, the space to nearest 81$P_{3/2}$ state is 1.3124~GHz, corresponding dipole matrix element 5649.1~$ea_0$ with $e$ electron charge and  $a_0$ Bohr radius. Therefore, 80$D_{5/2}$ Rydberg state displays strong dipole interactions with 81$P_{3/2}$, resulting to the state transfer of 80$D_{5/2}$ $\to$ 81$P_{3/2}${\color{red}~\cite{Hao_Observation_2021}}. This transformation leads to the decay of 80$D_{5/2}$ Rydberg atoms and further decreases of  EIT transmission. In order to verify this conjecture, we carry out more test that is performed in an additional MOT [not shown in Fig.~1(b)], in which Rydberg atoms is detected with a state select field ionization detection. The temperature of the atomic cloud is almost same with the main setup, and the peak density of atomic cloud is $8.0 \times 10^{10} cm^{-3}$, which is comparable with the main setup of $1.0 \times 10^{11} cm^{-3}$. In addition, we make the probe and coupling Rabi frequency on the almost same to that of the main setup so that the Rydberg population is comparable with that of the EIT in the main apparatus. Therefore, we obtain the similar EIT spectra and field ionization signal simultaneously in the test setup. The details of the setup can be seen in our previous work~\cite{Han2018, Hao_Observation_2021}. In the test experiments, after switching off the MOT beams, we apply a two-photon excitation pulse with duration 4~$\mu$s for preparing 80$D_{5/2}$ Rydberg atoms, an optional interaction time $t_{\rm INT}$ before the ionization detection allows us to study the decay of 80$D_{5/2}$ Rydberg and state transformation.

%(b) Measurements of the 80$D_{5/2}$ and 81$P_{3/2}$ states as a function of the interaction time $t_{\rm INT}$. Initially prepared 80$D_{5/2}$ Rydberg atoms transfer to nearby 81$P_{3/2}$
%state during $t_{\rm INT}$ due to the strong resonant dipole interaction between 80$D_{5/2}$ and 81$P_{3/2}$ atoms. The solid lines show the exponential fittings with the characteristic
%time $\tau_{0(81P_{3/2})}$ = 3.05 $\pm$ 1.23 $\mu$s and $\tau_{0(80D_{5/2})}$ = 3.15 $\pm$ 0.38 $\mu$s.

In Fig.~5(a), we present the time of flight (TOF) signals for laser excitation to 80$D_{5/2}$ state. Initially, atoms are populated in 80$D_{5/2}$ state (see the black curve) in the TOF signal. Due to the resonant dipole interaction,  atoms in 80$D_{5/2}$ state in light  pink shaded region  partly transfer to nearby 81$P_{3/2}$ state in the gray shaded region [see the red curve for $t_{\rm INT}$ = 3~$\mu$s].
With further increasing $t_{\rm INT}$, most of 80$D_{5/2}$ state Rydberg atoms transfer to 81$P_{3/2}$ state (see the blue curve for $t_{\rm INT}$ = 6~$\mu$s).  The population in state 81$P_{3/2}$ reaches maximal at $t_{\rm INT}$ = 6~$\mu$s, and then decays to other states.
For better understanding of the transfer process, we make a series of measurements for different $t_{\rm INT}$. Figure 5(b) displays the measured
80$D_{5/2}$ as a function of $t_{\rm INT}$. It is clear to observe the fast decay process on 80$D_{5/2}$ state within $8 {\rm \mu s}$.
To obtain the decay characteristic of 80$D_{5/2}$ state, we fit the experimental data using exponential function [see the solid line of Fig.~5(b)]. The fitted results show decay time $\tau_{0(80D)}$ = 3.15 $\pm$ 0.38 $\mu$s is close to the decay of the EIT transmission rate $\tau_{0(EIT)}$. Therefore, we conclude that the dipole interaction induced the state transfer may be the reason that leads to the decay of 80$D_{5/2}$ Rydberg three-level EIT transmission.

\section{Conclusion}

We have presented Rydberg EIT spectra in a cascade three-level scheme involving 80$D_{5/2}$ state of Cs atoms. Rydberg EIT spectrum shows strong dependence on the probe incident photon number. The optical transmission displays decay behavior with time. An optical depth of the medium is defined to characterize the transmission. We have shown that the optical depth displays linear increase with the time at onset for fixed probe $R_{\rm in}$.  We have further obtained the dephasing rate $\gamma_{\rm OD}$ by redefining OD=$\gamma_{\rm OD}t$. The dephasing rate linearly increases with weak probe field and then saturates for large $R_{\rm in}$. We have shown that the dephasing mechanism is mainly attributed to the strong dipole-dipole interactions, which lead to strong population decay on $|nD_{5/2}\rangle$ state. The experimental setting provides a platform to explore quantum nonlinear optics and quantum information processing, and creates metastable state in quantum many-body systems.

\end{document}